\documentclass{WileyMSP-template}

\begin{document}

% XX Die nächsten beiden Zeilen ein/aus kommentieren für Wiley/ArXiv Submission 
%\pagestyle{fancy}
%\rhead{\includegraphics[width=2.5cm]{vch-logo.png}}

\title{Review on coherent quantum emitters in hexagonal boron nitride}
%\title{Coherent quantum emitters in hexagonal boron nitride}

\maketitle

% Author: Please give full first and last names for authors and include * after the name of all corresponding authors

\author{Alexander Kubanek*}
%\author{Author One}
%\author{Author Two}

% Dedication

% \dedication{Optional dedication here. If no dedication is required, please leave blank}

% Affiliations: Please provide adacemic titles (Prof. or Dr.) for all authors where applicable, and include an institutional email address for all corresponding authors
\begin{affiliations}
Prof. Dr. Alexander Kubanek\\
Ulm University\\
Albert-Einstein-Allee 11\\
89081 Ulm\\
Germany\\
Email Address: alexander.kubanek@uni-ulm.de\\

\end{affiliations}

% Keywords: Please provide a minimum of three and a maximum of seven keywords, separated by commas


% Abstract should be written in the present tense and impersonal style (i.e., avoid we), and be at most 200 words long
\begin{abstract}
Hexagonal boron nitride is an emerging two-dimensional material with far-reaching applications in fields like nanophotonics or nanomechanics. Its layered architecture plays a key role for new materials such as Van der Waals heterostructures. The layered structure has also unique implications for hosted, optically active defect centers. A very special type of defect center arises from the possibility to host mechanically isolated orbitals localized between the layers. The resulting absence of coupling to low-frequency acoustic phonons turns out to be the essential element to protect the coherence of optical transitions from mechanical interactions with the environment. Consequently, the spectral transition linewidth remains unusually narrow even at room temperature, thus paving a new way towards coherent quantum optics under ambient conditions. In this review, I summarize the state-of-the-art of defect centers in hexagonal boron nitride with a focus on optically coherent defect centers. I discuss the current understanding of the defect centers, remaining questions and potential research directions to overcome pervasive challenges. The field is put into a broad perspective with impact on quantum technology such as quantum optics, quantum photonics as well as spin optomechanics. 

\end{abstract}

% Text: Please use section headings and subheadings as specified below. For communications, all section headings apart from Experimental Section should be removed
% Please make the first reference to a display item bold: \textbf{Figure 1}
% Do not abbreviate Figure, Equation, etc.; display items are always singular, i.e., Figure 1 and 2.
% Equations are always singular, i.e., Equation 1 and 2, and should be inserted using the {equation} environment, not as graphics
% Please do not use footnotes in the text, additional information can be added to the Reference list.

\section{Introduction}

\begin{figure}
  \centering
  \includegraphics[width=10 cm]{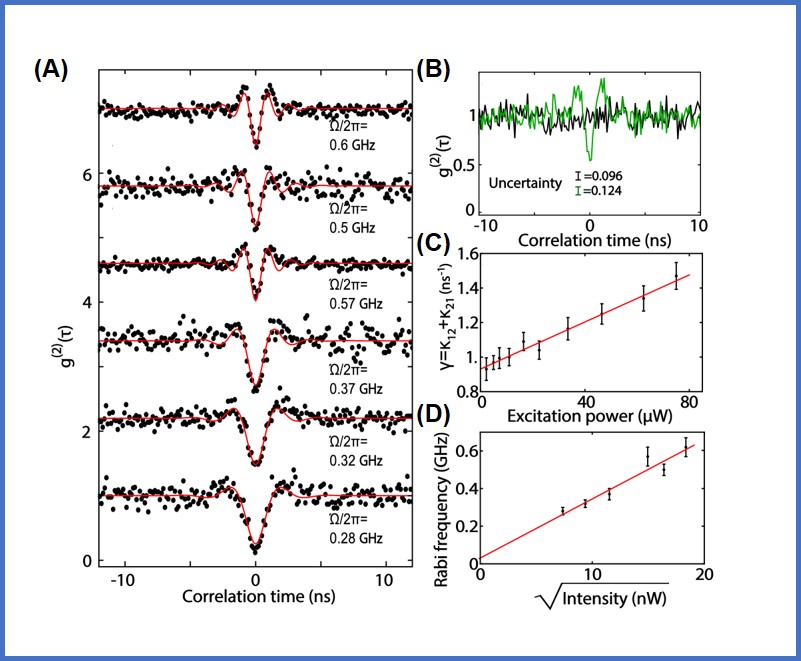}
  \caption{\textbf{The first signature of coherent Rabi oscillations.}  (A) Rabi oscillations are inferred from second-order photon correlations for strong, resonant driving and compared with a white light source for reference (B). The spontaneous decay rate (C) and estimated Rabi frequency (D) decay linear with the excitation power and, respectively, with the square root of the drive intensity. Reprinted with permission from \cite{konthasingheRabiOscillationsResonance2019a}. Copyright 2019, Optica Publishing Group under the terms of the Open Access Publishing Agreement. DOI: /10.1364/OPTICA.6.000542}
  \label{fig:Rabi}
\end{figure}

Optically active defect centers are lattice imperfections on the atomic level that interact with optical fields. Such defect centers can be found in a variety of solid-state materials. Defect centers in wide band-gap materials are an important element for scalable quantum devices with strong light-matter interaction and pronounced optical non-linearities \cite{awschalomQuantumTechnologiesOptically2018a, atatureMaterialPlatformsSpinbased2018a}. 
Further impact arises when the defect center's optical transition is coupled to the defect center's intrinsic spin  or to an nearby spin in the solid state environment. Color centers in diamond \cite{dohertyNitrogenvacancyColourCentre2013} or silicon carbide \cite{christleIsolatedElectronSpins2015} possess  long spin coherence times \cite{maurerRoomTemperatureQuantumBit2012a} and spin manipulation with high-fidelity \cite{robledoHighfidelityProjectiveReadout2011}. Defect centers in two-dimensional or layered materials have shown promising photophysical properties \cite{aharonovichSolidstateSinglephotonEmitters2016a} and stand out by unique capabilities arising from the reduced thickness of just a few atomically thin layers. New opportunities emerge, for example, for optical integration into quantum photonic and plasmonic devices \cite{aharonovichQuantumEmittersTwo2017} or cavity optomechanics \cite{aspelmeyerCavityOptomechanics2014}. The defect centers sensitivity to optical and microwave fields as well as electrical, magnetic or strain fields make them promising candidates for quantum sensing \cite{mazeNanoscaleMagneticSensing2008} with the potential for sensing on sub-nanometer scale \cite{gottschollSpinDefectsHBN2021, hayeeRevealingMultipleClasses2020}.
 Quantum sensing is among the most developed applications, especially, because coherent optical transitions are often not required relaxing the technical overhead. In contrast, spin-photon interfaces for entanglement distribution almost exclusively rely on coherent optical transitions. However, optical transitions of quantum emitters in solids are very sensitive to interactions with the environment, which typically lead to rapid emitter dephasing. Sample cooling is therefore essential in order to suppress in particular phonon-mediated interactions with the environment. Hence, experiments are currently performed in cryogenic environments.
In order to realize applications such as quantum networks \cite{wehnerQuantumInternetVision2018a} in a practical way an efficient distribution of quantum entanglement is indispensable. Quantum photonics enables the efficient distribution of quantum entanglement \cite{sangouardLongdistanceEntanglementDistribution2007} and is a crucial element for a variety of protocols in quantum information processing. Quantum communication enables secure quantum-key distribution with bright sources of linearly polarized single photons \cite{obrienPhotonicQuantumTechnologies2009, lounisSinglephotonSources2005}. Heralded excitation mechanisms enable deterministic photon-pair generation based on spontaneous parametric down-conversion sources mediated by the presence of a single two-level system \cite{saraviAtommediatedSpontaneousParametric2017}. \\
Overcoming the need for cryogenic temperatures is an ubiquitous goal. Proposals on solid-state emitters embedded in cavities with ultrasmall mode volume promise efficient indistinguishable single photon generation at room temperature \cite{weinFeasibilityEfficientRoomtemperature2018}. Another, very recent discovery shows that the layered structure of a two-dimensional material could play a key role to overcome the requirements for sample cooling. The orbitals of hosted defect centers could be oriented in a mechanically isolated way such that optical transitions remain decoupled from phonon interactions of relevant phonon modes \cite{hoeseMechanicalDecouplingQuantum2020}. Although the exact orbital structure that enables the isolation is still unclear. Ultimately, such mechanically isolated defect centers could enable coherent light-matter interaction where the emitter's optical coherence is protected by mechanical isolation instead of sample cooling. Therefore, solid-state quantum optics under ambient conditions could become reality turning into an applied technology.\\
\textbf{H}exagonal \textbf{b}oron \textbf{n}itride (h-BN) is an emerging platform which stands out with its large band gap of 6.08 eV \cite{cassaboisHexagonalBoronNitride2016}. Developments on the h-BN material processing enabled single-layer h-BN growth of high quality \cite{watanabeDirectbandgapPropertiesEvidence2004}. The advances on the material quality paved the way for applications in optoelectronics and in quantum photonics \cite{caldwellPhotonicsHexagonalBoron2019} with the potential to operate quantum light sources at room temperature \cite{tranRobustMulticolorSingle2016, kianiniaRobustSolidStateQuantum2017}. While the single photon emission of defect centers in h-BN was extensively studied, coherent transitions remain to a large extent unexplored. Not long ago, the first Rabi oscillations were reported by laser-induced coherent population cycling of an optical transition \cite{konthasingheRabiOscillationsResonance2019a} marking the first step towards coherent optical control at cryogenic temperatures. The Rabi oscillations were inferred from  photon statistics by recording the second-order correlation function $g_2(\tau)$, as summarized in Figure \ref{fig:Rabi}. Another measure for the optical coherence is the spectral linewidth of an emitters transition. A linewidth given by the excited states lifetime signals a coherent optical transition in the so-called \textbf{F}ourier \textbf{t}ransform \textbf{l}imit (FTL) \cite{dietrichObservationFourierTransform2018a}. \\

This review focuses on defect centers in h-BN and, especially, on a specific class of defect centers with coherent optical transitions at room temperature. The review illuminates the current understanding of the defect centers, discusses actual limitations and pervasive challenges and outlines potential future directions. In particular, I discuss potential applications in quantum photonics and spin optomechanics.

\section{The role of the layered architecture of h-BN}

The two-dimensional structure of h-BN is similar to the structure of graphene and has far-reaching implications on the usability of h-BN. In this section, I focus on the role of the layered architecture distinguishing between monolayers and multilayers of h-BN. 

\subsection{Monolayers of two-dimensional h-BN}

\begin{figure}
  \centering
  \includegraphics[width=16 cm]{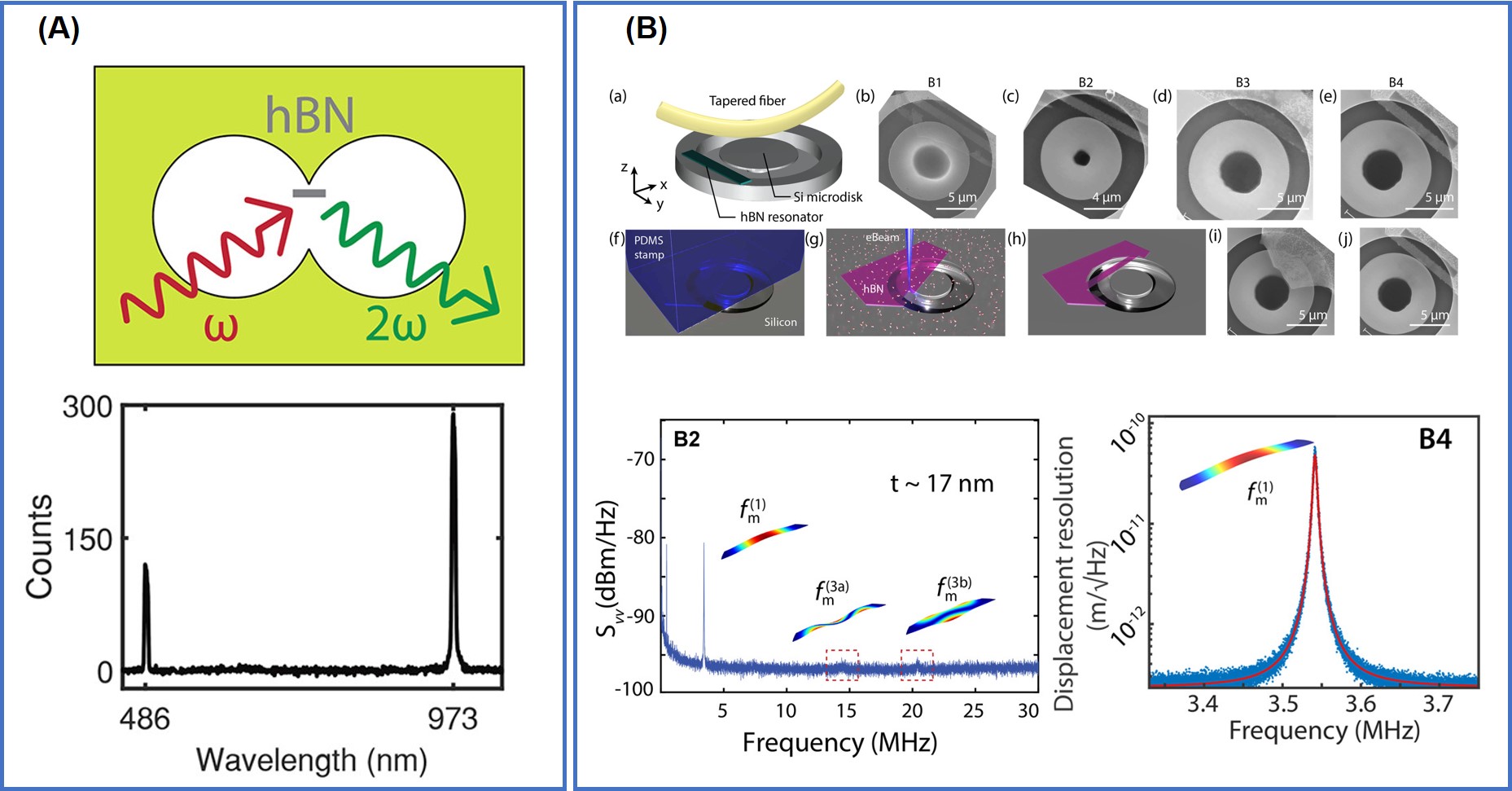}
  \caption{\textbf{Applications of h-BN in nonlinear optics and optomechanics.} (A) The strong optical nonlinearity originating from subwavelength, two-dimensional h-BN using plasmonic optical tweezer. The local field and local density of optical states is enhanced by a double nanohole plasmonic structure (upper panel). The resulting nonlinear optical response is strongly enhanced with up to two orders of magnitude higher power in the second harmonic generation as compared to competing nonlinear materials (lower panel). (B) The first cavity optomechanical system with incorporated h-BN. The frequencies of thermally driven resonances range from 1 to 23 MHz with Q-factors up to 1100 at 23 MHz. The Lorentzian fit yields a mechanical Q-value of 678 at $f_m^{(1)} \approx 3,45 \, MHz$ reaching a maximum sensitivity of 0.16 pm / $\sqrt{Hz}$ to the motion of a h-BN nanobeam. 
(A) Reprinted with permission from \cite{hajisalemSingleNanoflakeHexagonal2021}. Copyright 2021, American Chemical Society. DOI: 10.1021/acsphotonics.1c00525.
(B) Reprinted with permission from \cite{shandilyaHexagonalBoronNitride2019}. Copyright 2019, American Chemical Society. DOI: 10.1021/acs.nanolett.8b04956.}
  \label{fig:Opto}
\end{figure}

The strong nonlinear optical response associated with two-dimensional materials facilitates efficient second harmonic generation from sub-wavelength, two-dimensional nanoflakes. The two-dimensionality entails that the nanoflakes can be positioned with high precision, for example in regions of highest fields in plasmonic devices, which results in the generation of a nonlinear optical response with highest efficiency even for low-power, CW-laser excitation \cite{hajisalemSingleNanoflakeHexagonal2021} as depicted in Figure \ref{fig:Opto} (A). The fact that h-BN is a two-dimensional material, conditions that the incorporated defect centers are located directly at the host surface, which makes it an ideal platform for evanescent integration into plasmonic and photonic devices \cite{prosciaMicrocavitycoupledEmittersHexagonal2020}. Besides applications originating from the evanescent coupling, the h-BN-material enables direct structuring of photonics devices into the host material \cite{kimPhotonicCrystalCavities2018a}. The direct proximity between defect center and host surface furthermore accommodates applications in quantum sensing with ultra-short distance between the atomic sensor and the target \cite{gottschollSpinDefectsHBN2021}. The short-range interaction leads to nanoscale quantum sensing with enhanced sensitivity of h-BN spin defects for magnetic field detection. Furthermore, one of the most prominent applications for monolayer h-BN are van der Waals heterostructures composed of individual monolayers \cite{geimVanWaalsHeterostructures2013, liuVanWaalsIntegration2019}. Returning to nonlinear optics, van der Waals heterostructures can be stacked with arbitrary angles between the layers. Controlling the inversion symmetry at the crystal interfaces can be used to introduce broken symmetries that induce a strong nonlinear optical response. Such tunable second harmonic generation was demonstrated with twisted stacks in h-BN homostructures by controlling the twist angle and the underlying moire interface \cite{yaoEnhancedTunableSecond2021}.

\subsection{Multilayers of two-dimensional h-BN}

\begin{figure}
  \centering
  \includegraphics[width=10 cm]{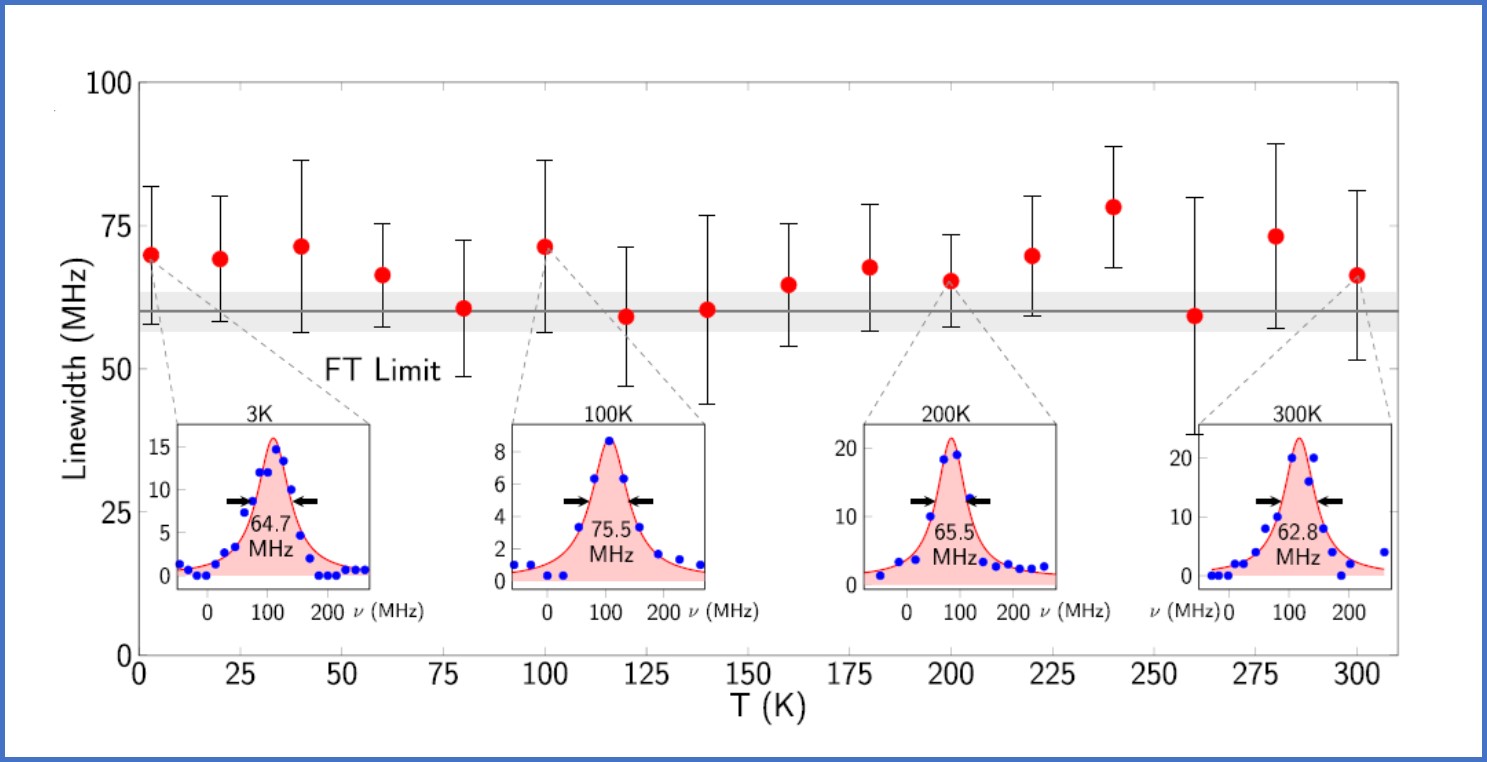}
  \caption{\textbf{The persistence of narrow optical transitions up to room temperature.} Photoluminescence excitation (PLE) scans uncover a homogeneous linewidth of about 60 MHz within the FTL. The narrow linewidth persists over an unusual temperature range from 3 K to room temperature. Reprinted with permission from \cite{dietrichSolidstateSinglePhoton2020a}. Copyright 2020, by the American Physical Society. DOI: 10.1103/PhysRevB.101.081401.}
  \label{fig:LW}
\end{figure}

Also multilayer h-BN and nanocrystals, which I refer to as quasi-bulk when the layers stack up to many nm-thickness, obey unique capabilities arising from the layered architecture. For example, integrated optomechanical circuits is an emerging technology based on quasi-bulk h-BN. Studying the mechanical properties of h-BN by means of cavity optomechanics for two-dimensional materials is challenging since the observation of the mechanical motion is complicated at nanometer scale. Decreasing the mechanical resonator mass in the context of nanomechanics enhances the performance. The low-mass of the devices are associated with large zero point motion compared to conventional dielectric films, which are used in quantum optomechanics. As a hyperbolic material, h-BN supports the propagation of phonon polaritons capable of strong optical confinement in the sub-wavelength regime. A cavity optomechanical system with incorporated h-BN was realized by positioning h-BN beams in the optical near-field of silicon microdisk cavities that enable the observation of thermally driven mechanical resonances in the MHz frequency range with large mechanical quality factors above 1000 at room temperature and in high vacuum \cite{shandilyaHexagonalBoronNitride2019}, as shown in Figure \ref{fig:Opto} (B).\\
A very unique role of the layered architecture for quasi-bulk h-BN becomes apparent when looking at the hosted quantum emitters. First of all, the large band-gap energy supports a large variety of defect centers. In addition, the layered structure enables to host quantum emitters with their orbitals located between the layers. Consequently, the orbitals could be isolated from direct interactions with the solid-state environment. The quantum emitters could, in principle, include individual atoms or molecules, which are trapped between the layers as well as defect centers, which are distorted out-of-plane. As it turned out, the out-of-plane displacement of the defect center orbitals leads to a set of unique properties, in particular, FTL spectral lines without any dephasing or spectral diffusion under resonant excitation on the timescale of the resonant laser scan. Typical on-times of the emitters fluorescence under resonant excitation are in the milliseconds range with maximum reported timescales of many seconds \cite{dietrichObservationFourierTransform2018a}. Uniquely, these narrow optical transitions remain narrow up to room temperature, as depicted in Figure \ref{fig:LW}. In the following, I discuss the current understanding of these emitters. 

\section{The large family of defect centers in h-BN}

The h-BN insulator has a wide band-gap energy of 6eV, which is among the largest of all two-dimensional materials. Consequently, h-BN has a large transparency window from visible to infrared wavelengths and is capable to host many optically active defect centers with optical transition wavelength distributed from 580 nm to 800 nm  \cite{jungwirthTemperatureDependenceWavelength2016, tranRobustMulticolorSingle2016, dietrichObservationFourierTransform2018a}. Unveiling the defects composition and local structure is an ongoing research effort with a close link between experiment and theory  \cite{auburgerInitioIdentificationParamagnetic2021, hayeeRevealingMultipleClasses2020, hayeeRevealingMultipleClasses2020a, westonNativePointDefects2018, tawfikFirstprinciplesInvestigationQuantum2017, choiEngineeringLocalizationQuantum2016}. In the following I give a small, exemplary insight into the discussion on the microscopic origin of specific single-photon transitions in h-BN  without the claim of a complete overview. Examples include boron dangling bonds as source for single photon emission at 2.06 eV \cite{turianskyDanglingBondsHexagonal2019} which could explain the accumulation of emitters localized near crystal edges or grain boundaries \cite{chejanovskyStructuralAttributesPhotodynamics2016, xuSinglePhotonEmission2018}. UV emission at 4.08 eV
was traced back to optically active, pentagon-hexagon Stone-Wales defects preferentially present in poly-crystalline h-BN \cite{hamdiStoneWalesDefects2020}. Emitters with transition energies between 2.0 and 2.2 eV were also found to originate from carbon-related defects \cite{mendelsonIdentifyingCarbonSource2021}. The atomic structure and radiative properties for more than 20 native defects as well as carbon and oxygen impurities in h-BN was investigated by ab initio density functional theory and Bethe-Salpeter equation calculations indicating exciton energies from 0.3 to 4 eV with radiative lifetimes between ns to ms \cite{gaoRadiativePropertiesQuantum2021}. Native $V_NN_B$ defects were identified as most likely candidates by Bayesian statistical analysis. Carbon dimer defects $C_BC_N$ generated attention as candidates for spectrally narrow, single photon emission in the UV with a ZPL in the 4.1 eV band (at 4.08 eV) with modest electron-phonon coupling, short lifetime of about 1.2 ns and a potential quantum efficiency close to 1 resulting in a high photon yield \cite{mackoit-sinkevicieneCarbonDimerDefect2019, duOriginDeeplevelImpurity2015, vokhmintsevElectronphononInteractionsSubband2019, bourrellierBrightUVSingle2016, meuretPhotonBunchingCathodoluminescence2015a, uddinProbingCarbonImpurities2017}.\\
A key experimental input to shed more light on the defects compositions could be the deterministic creation of emitters in h-BN with known atomic constituents. Defect centers were created by focused ion beam (FIB) milling arrays of holes into h-BN monolayers, where single quantum emitters form at the edge of the holes with a yield of up to 31 \% \cite{zieglerDeterministicQuantumEmitter2019}. Alternatively, single photon emitters were created by combining nanoscale strain engineering and charge trapping, where the emitter activation results from carrier trapping in deformation of potential wells at positions where the h-BN film reaches the highest curvature \cite{prosciaNeardeterministicActivationRoomtemperature2018a}.
Other creation methods include electron irradiation, high temperature annealing \cite{tranRobustMulticolorSingle2016}, ion implantation \cite{choiEngineeringLocalizationQuantum2016}, chemical vapour deposition (CVD) \cite{mendelsonEngineeringTuningQuantum2019, sternSpectrallyResolvedPhotodynamics2019, abidiSelectiveDefectFormation2019} as well as plasma treatment \cite{voglFabricationDeterministicTransfer2018}. Optically addressable spin ensembles have been created by femtosecond laser writing \cite{gaoFemtosecondLaserWriting2021}. \\
It is beyond the scope of this review to discuss the details of each individual defect center and its properties. Instead, I discuss more general remarks on photophysical and spin properties and then focus on defect center with coherent optical transitions. 
  
\subsection{Photophysical properties}

Defect centers in h-BN are among the brightest single photon sources \cite{martinezEfficientSinglePhoton2016, grossoTunableHighpurityRoom2017a} with large Debye-Waller factor $>80$ \% \cite{tranQuantumEmissionHexagonal2016},  high polarization contrast \cite{exarhosOpticalSignaturesQuantum2017} and high robustness \cite{kianiniaRobustSolidStateQuantum2017}. The three-dimensional dipole orientation of individual defect centers has been analyzed \cite{nikolayDirectMeasurementQuantum2019a, takashimaDeterminationDipoleOrientation2020}. High-purity single photon emission was demonstrated with spectral tunability over 6 meV by strain control \cite{grossoTunableHighpurityRoom2017a}
as well as by applied electric fields \cite{nohStarkTuningSinglePhoton2018a, nikolayVeryLargeReversible2019}. Quantum efficiencies were reported as high as $87$ \% \cite{nikolayDirectMeasurementQuantum2019a}. At room temperature, the defect centers show sharp optical transitions with linewidths of a few nm in the visible wavelength range \cite{tranQuantumEmissionHexagonal2016, grossoTunableHighpurityRoom2017a, voglFabricationDeterministicTransfer2018}. 
At cryogenic temperatures, narrowband transitions with sharp zero-phonon lines (ZPL) were reported \cite{jungwirthTemperatureDependenceWavelength2016, sontheimerPhotodynamicsQuantumEmitters2017, duongFacileProductionHexagonal2019, exarhosOpticalSignaturesQuantum2017, khatriOpticalGatingPhotoluminescence2020, shotanPhotoinducedModificationSinglePhoton2016} with linewidths down to about 60 MHz \cite{dietrichObservationFourierTransform2018a}. Emitters with such narrow spectral transitions reaching the FTL are subject to a detailed discussion in the following section. More advanced excitation and absorption schemes, such as two-photon absorption \cite{schellNonlinearExcitationQuantum2016} and anti-Stokes excitation \cite{tranSuppressionSpectralDiffusion2019}, has been reported.

\subsection{Thoughts on spin qubits in h-BN}

Defect centers in h-BN with addressable spin at room temperature were discovered \cite{mendelsonIdentifyingCarbonSource2021a, chejanovskySinglespinResonanceVan2021a, sternRoomtemperatureOpticallyDetected} demonstrating optically detected magnetic resonance (ODMR) \cite{gottschollInitializationReadoutIntrinsic2020, chejanovskySinglespinResonanceVan2021}.
Carbon trimer defects, in particular the $C_2C_N$-trimer, were investigated in detail by ab inito and group theory finding. For example, several radiative transitions together with spin-orbit and spin-spin assisted non-radiative transitions and a possible ODMR signal with a contrast of $\approx 1$ \% were found \cite{golamiTextitAbInitioGroup2021}. Defects involving a single negatively-charged boron vacancy ($V_B^-$) emitting preferentially at 800 nm were discovered as spin defects with a spin S=1 \cite{gottschollInitializationReadoutIntrinsic2020} demonstrating ODMR \cite{abdiColorCentersHexagonal2018, ivadyInitioTheoryNegatively2020, reimersPhotoluminescencePhotophysicsPhotochemistry2020a, gottschollRoomTemperatureCoherent} 
and rabi oscillations \cite{liuRabiOscillationTextB2021}. Recently, record-high ODMR contrast of 46 $\%$ was demonstrated for shallow $V_B^-$ at room temperature making use of surface plasmons of a gold film microwave waveguide \cite{gaoHighContrastPlasmonicEnhancedShallow2021}. The resulting CW ODMR sensitivity reaches $8 \, \mu T / \sqrt{Hz}$ and hints towards potential applications in nanoscale sensing. An open question remains on the spin-purity of h-BN with its spin-containing lattice components. Can h-BN reach a spin-purity comparable to competing platforms such as diamond, which can be engineered up to an extreme level of isotopic purity? Spin noise in h-BN could potentially limit the achievable spin coherence therefore limiting spin-based applications, for example in quantum sensing. However, lacking spin coherence could be compensated by the proximity of the defect center to the surface. Sensing at Angstrom-distances could reach a high level of sensitivity and resolution, even for shorter coherence times. As an example, $V_B^-$ was used as atomic scale sensors for temperature, magnetic fields and applied pressure \cite{gottschollSpinDefectsHBN2021}. Demonstrating sensing with only three layers of h-BN, where the intermediate layer hosts the $V_B^-$-center, results in a minimum distance of $\approx 0.33$ nm.
 
\section{The special class of mechanically isolated defect centers}

I now turn the discussion to mechanically isolated defect centers in h-BN. Some defect centers showed optical transitions with very narrow linewidth within the FTL in photoluminescence excitation (PLE) spectroscopy \cite{dietrichObservationFourierTransform2018a}. Astonishingly, these optical transitions remain within the FTL over an unusual temperature range up to room temperature. First investigations suggested a mechanical isolation mechanism as origin for the persistence of the spectrally narrow lines. I introduce the current understanding of this special class of emitters, discuss how they can be distinguished from other defect centers and comment on upcoming challenges and potential impact.

\subsection{Electron-phonon coupling}

\begin{figure}
  \centering
  \includegraphics[width=17 cm]{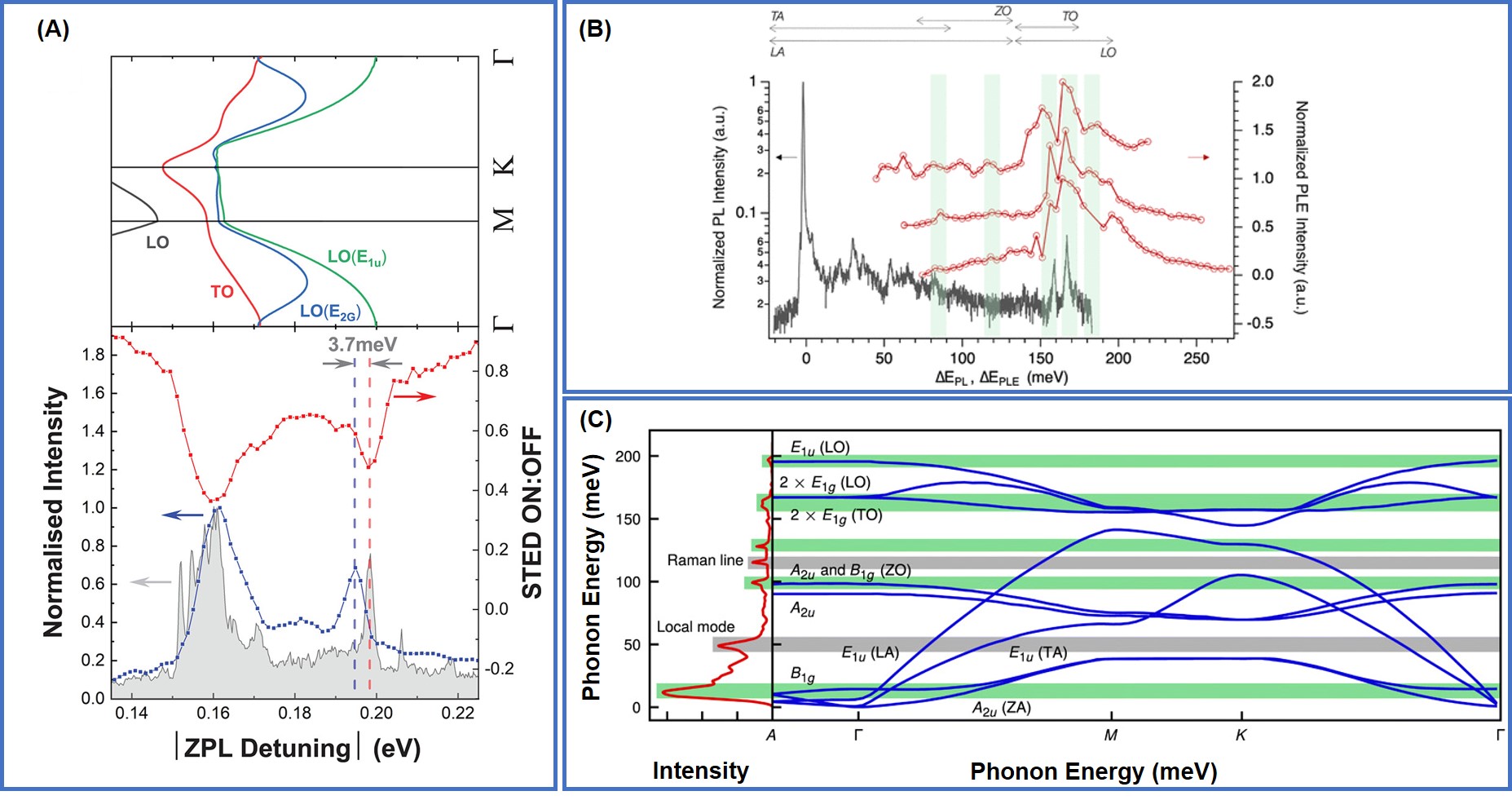}
  \caption{\textbf{The electron-phonon coupling to bulk and local phonon modes in h-BN.} (A) The optical phonon dispersion for bulk h-BN was calculated in reference \cite{serranoVibrationalPropertiesHexagonal2007} (upper panel). A comparison with the PSB spectra measured with PL, PLE as well as STED enables to assigned the specific bulk phonon modes (lower panel). 
(B) Resonances in the emission spectrum (black data) and PLE spectra (red data) are associated with different phonon modes marked with green areas. 
(C) A comparison of the PSB emission in PL (left panel) with the PSB-decomposition (right panel) furthermore shows individual features which are not reproduced from bulk modes and are associate with local phonon modes instead.
(A) Reproduced under terms of the CC-BY license from \cite{maleinStimulatedEmissionDepletion2021}. Copyright 2021, American Chemical Society. DOI:10.1021/acsphotonics.0c01917.   
(B) reproduced with permission from \cite{grossoLowTemperatureElectronPhonon2020}. Copyright 2020, American Chemical Society. DOI: 10.1021/acsphotonics.9b01789.
(C) reproduced with permission from \cite{hoeseMechanicalDecouplingQuantum2020}. Copyright 2020, Science Advances, some rights reserved; exclusive licensee AAAS. Distributed under a CC BY-NC 4.0 License. DOI: 10.1126/sciadv.aba6038.}
  \label{fig:ep}
\end{figure}

Investigating the electron-phonon coupling is a crucial cornerstone for understanding the defect centers and their interactions with the environment. The fingerprint of the electron-phonon coupling can be used to draw conclusions on the emitters composition and its local solid-state surrounding. The phonon sideband (PSB) emission harbors the information on the coupling to phonons. A detailed study shows that the observed PSB can be explained by coupling to bulk longitudinal optical (LO) phonons, and the asymmetric ZPL by additional longitudinal acoustic phonons (LA) as well as local modes \cite{wiggerPhononassistedEmissionAbsorption2019}. The electron-optical-phonon sideband of the excited state can be investigated by PLE measurements, where resonant driving of the phonon resonances discloses the coupling of individual modes  \cite{hoeseMechanicalDecouplingQuantum2020, grossoLowTemperatureElectronPhonon2020}. Investigations on the emitter's coupling efficiency to different phonon modes by means of PLE measurements demonstrates that the excitation mediated by the absorption of one in-plane optical phonon increases the emitter absorption probability by a factor of 10 as compared to that mediated by acoustic or out-of-plane optical phonons \cite{grossoLowTemperatureElectronPhonon2020}. The first ZA (out-of-plane acoustic) phonon mode 8 meV detuned from the ZPL can be efficiently excited addressing the same excited state of the ZPL with the same polarization contrast. Complementary, the electron-optical-phonon sideband of the ground state can be mapped by stimulated emission depletion (STED) spectroscopy \cite{maleinStimulatedEmissionDepletion2021}. A red-shift in the LO-phonon mode with $E_{1u}$-symmetry was observed for color centers emitting around 563 nm (2.2 eV), which is consistent with the $(1)^4 B_1 \rightarrow (1)^4 A_2$ transition in the $V_BC_N^-$-defect, tracing back the phononic fingerprint to carbon-related defects in this case. As an example, Figure \ref{fig:ep}(A) depicts the calculated optical phonon dispersion for bulk h-BN, taken from \cite{serranoVibrationalPropertiesHexagonal2007}. A comparison to the measured photoluminescence (PL), PLE as well as STED spectra enables to assign the individual bulk modes.\\
The electron-phonon spectral density was extracted from the PL and PLE spectrum\cite{grossoLowTemperatureElectronPhonon2020} and by phonon sideband decomposition \cite{hoeseMechanicalDecouplingQuantum2020} and h-BN phonon modes can be associated to its resonances, as depicted in Figure \ref{fig:ep}(B) and Figure \ref{fig:ep}(C), respectively. Herewith, a one-phonon band model reproduces the PSB suggesting that strong vibronic interactions, which involve non-symmetric modes or anharmonic effects, such as Jahn-Teller, can be neglected. Comparing the one-phonon mode decomposition with the modes of pristine h-BN indicates that acoustic modes in the out-of-plane direction with out-of-plane wave vector at the edge of the Brillouin zone (A-point) as well as out-of-plane optical modes at the $\Gamma$-point obey a high density of modes and a strong coupling to the defect. In contrast, in-plane displacements between neighboring unit cells of h-BN (longitudinal acoustic $E_{1u}$ mode at the K-point,  multiple transverse modes and longitudinal optical $E_{1u}$-mode at the $\Gamma$-point) only weakly couple to the defect. This observation suggests, that the defect centers only couple strongly to modes with out-of-plane displacements. The insensitivity to in-plane acoustic modes could be explained by orbitals of the defect centers, which extend out-of-plane. Furthermore, polarization-dependent PL-measurements suggest that phonons in the emission process do not alter the polarization contrast, which is consistent with the defect coupling to linear symmetric modes. Electromagnetic coupling between the optical transition and an optical mode results in significant electron-phonon coupling even for an emitter that is displaced out-of-plane. Accordingly, a large electron-phonon coupling strength for the optical phonon mode (165 meV detuned from the ZPL) was reported \cite{wiggerPhononassistedEmissionAbsorption2019} and efficient absorption via these modes was demonstrated. Again, the phonon-assisted excitation does not alter the polarization properties. \\
The investigated electron-phonon coupling can shed light on the mechanism that enables the narrow spectral linewidth of the optical transitions to persist up to room temperature and is discussed in the next section.

\subsection{Unique features of out-of-plane distorted defect centers}

\begin{figure}
  \centering
  \includegraphics[width=12 cm]{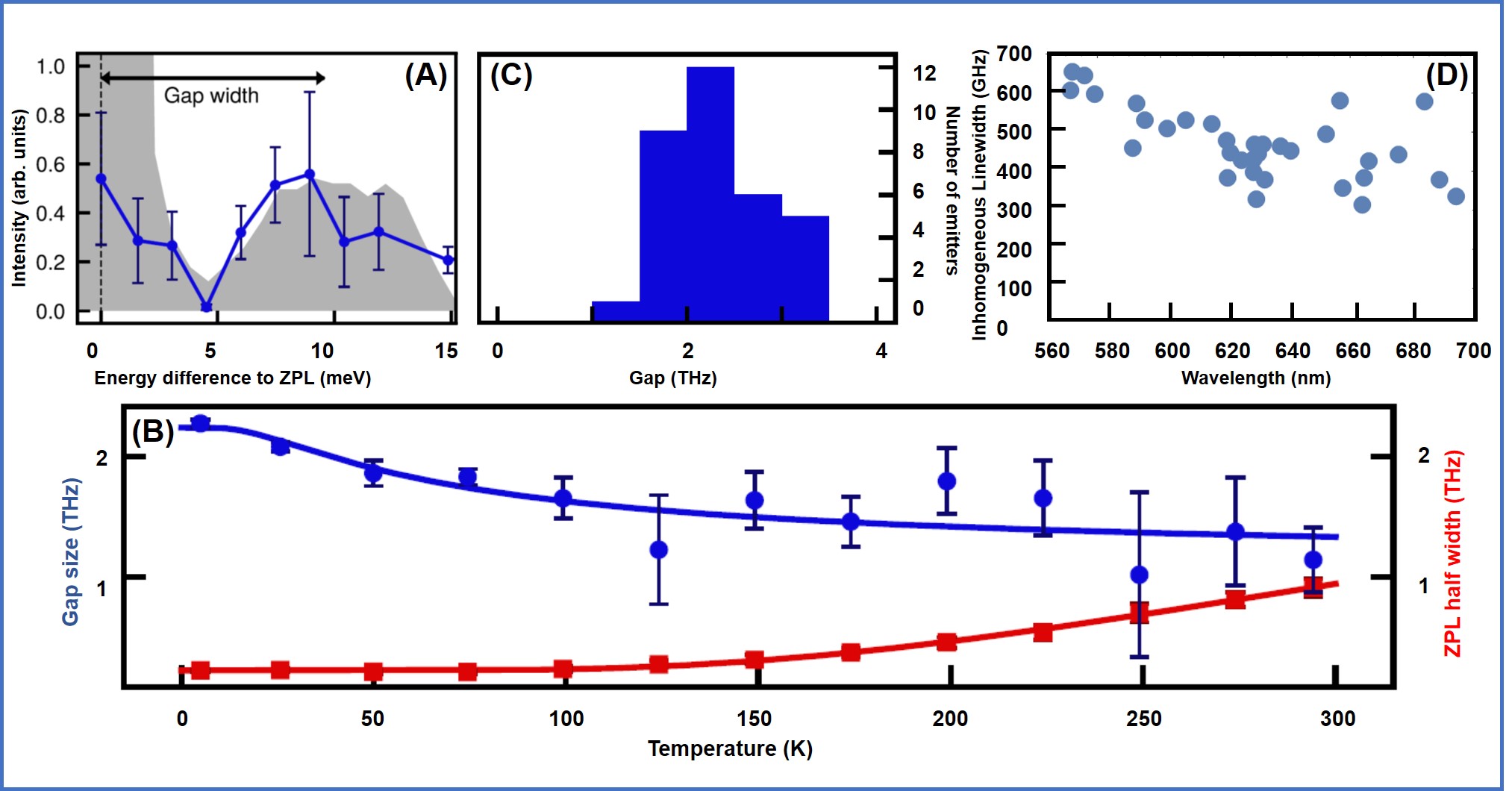}
  \caption{\textbf{The characteristic signature of mechanically decoupled defect centers.} (A) The characteristic emission (PL, grey shaded area) and excitation (PLE, blue data points) spectra show a low frequency gap about 5 meV detuned from the zero-phonon line (ZPL). (B) The temperature dependence shows a reduction in gap size with increasing temperature. But the gap size remains larger than the ZPL linewidth at room temperature. (C) The gap size is typically about 2 THz at 4 K confirmed by about 30 emitters. (D) The emitters with gaps in the low-frequency electron-phonon density of states were found with ZPL wavelengths between 560 and 700 nm. Besides the very narrow homogeneous linewidth, the inhomogeneous linewidths are broadened up to many hundreds of GHz. The spectral diffusion was accumulated over timescales of minutes. (A), (B)and (C) from Sci. Adv. 6, eaba6038 (2020), reproduced with permission \cite{hoeseMechanicalDecouplingQuantum2020}. Copyright 2020, Science Advances, some rights reserved; exclusive licensee AAAS. Distributed under a CC BY-NC 4.0 License.}
  \label{fig:Gap}
\end{figure}

A key questions remains: "How can mechanically isolated defect centers be distinguished from other defect centers in h-BN"? Paying attention to the PL-spectrum at low frequencies shows the clear signature of a gap between the ZPL and the acoustic phonon branches, shown as grey area in Figure \ref{fig:Gap}(A). Accordingly, the electron-phonon coupling strength has a minimum in that region (around 5 meV detuned from the ZPL) signaling the mechanical decoupling from in-plane phonon modes. The low electron-phonon coupling strength to in-plane acoustic phonons was reproduced by PLE-measurements as depicted by the blue data points in Figure \ref{fig:Gap}(A). The gap size between ZPL and first phonon resonance is determined to an energy of about 2 THz at 5K. The temperature dependence of the gap size in the electron-phonon spectral density is shown in Figure \ref{fig:Gap}(B). The low-frequency gap persists up to room temperature and explains the observed  narrow  lines  at  300 K. The gap size of 2 THz at 5K is reproduced by more than 30 emitters, shown in Figure \ref{fig:Gap}(C). The gap size narrows with increasing temperature while the ZPL linewidth increases, likely due to temperature-dependent distortion that modifies the electron-phonon couplings and phonon frequencies. However, the temperature dependence reveals that even at 300 K the gap size remains larger than the ZPL linewidth. Studying more than 30 emitters with a low-frequency gap in the electron-phonon density of states shows a distribution of transition wavelength across a wide range from 560 nm to 700 nm as depicted in Figure \ref{fig:Gap}(D). The broad PL linewidths, shown on the y-axis of Figure \ref{fig:Gap}(D), illustrates that on-going spectral diffusion is still dominating the spectrum under off-resonant excitation. The broad distribution of ZPL wavelengths could, in principle, be explained by one emitter type with a strong susceptibility to strain \cite{liGiantShiftStrain2020}. However, such strong strain susceptibility seems counterintuitive to the mechanical isolation mechanism. Alternatively, the large spread of ZPL wavelengths for mechanically isolated defect centers could suggest that the mechanical isolation mechanism is not specific to a particular type of defect center nor relies on a unique defect composition. It rather indicates that the isolation mechanism originates from a more fundamental process. Considering the layered architecture of h-BN, the mechanically decoupled emitters could be located between the layers or  their orbitals could be distorted out-of-plane. Consequently, the orbitals are less susceptible to in-plane phonon distortions and are therefore effectively decoupled from in-plane acoustic phonon modes. The exact feature of the orbitals which causes the mechanical isolation of specific optical transitions from their coupling to in-plane phonon modes needs further investigations. However, the intuitive picture of an out-of-plane distortion is supported by looking at the dipole emission directionality. For single defects in h-BN nanoflakes, which are not mechanically isolated, three-dimensional determination of the dipole orientation yielded, that the dipoles are oriented near the plane of the h-BN layers \cite{takashimaDeterminationDipoleOrientation2020}. Indeed, non-isolated emitters should have dipoles located in-plane with the h-BN layers and, consequently, predominantly emit photons perpendicular to the h-BN layers. In contrast, mechanically isolated defect centers with distorted orbitals require, that the dipole emission is non-orthogonal with respect to the h-BN layer. This was experimentally verified by correlating the tilt angle of the h-BN flakes with the photon emission directionality  of mechanically isolated defect centers \cite{hoeseMechanicalDecouplingQuantum2020}.

\section{Open questions, limitations and challenges}

\begin{figure}
  \centering
  \includegraphics[width=15 cm]{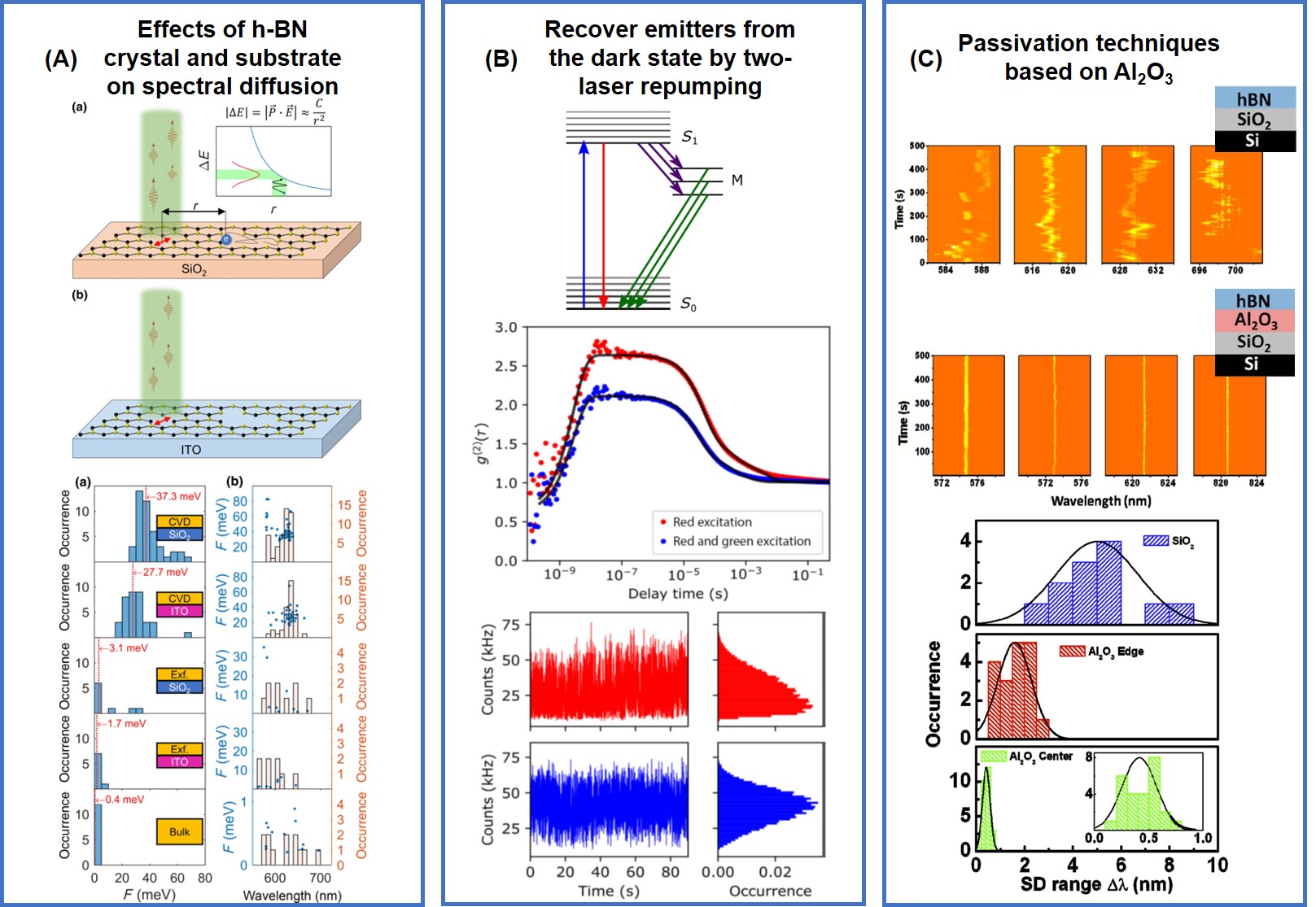}
  \caption{\textbf{Promising directions to improve spectral diffusion.} (A) Conducting substrates reduce the number of nearby charge traps, thereby suppressing the effect of spectral diffusion.  (B) A two-laser repumping scheme for resonant excitation. Repumping at 500 nm, recovers emitters from the dark state and improves the spectral stability. (C) Introducing a passivation layer of $Al_2O_3$ improves the spectral diffusion significantly yielding an improved diffusion linewidth by an order of magnitude.
(A) reprinted with permission from \cite{akbariTemperaturedependentSpectralEmission2021}. Copyright 2021, American Physical Society. DOI: 10.1103/PhysRevApplied.15.014036.
(B) reprinted with permission from \cite{whiteOpticalRepumpingResonantly2020}. Copyright 2020, American Physical Society. DOI: 10.1103/PhysRevApplied.14.044017.
(C) reprinted with permission from \cite{liNonmagneticQuantumEmitters2017a}. 
Copyright 2017, American Chemical Society. DOI: 10.1021/acsnano.7b00638.}
  \label{fig:Diffusion}
\end{figure}

Understanding the composition of defect centers in h-BN is crucial. A precise knowledge of the constituents and their properties is, for example, necessary for the deterministic creation of defect centers in the future. However, the mechanical decoupling mechanism seems to be quite universal as indicated by the large spread of ZPL wavelengths, depicted in Figure \ref{fig:Gap}(D). Therefore, the mechanical decoupling mechanism could be independent of the defects composition. Alternatively, strong electron-phonon coupling could lead to a large shift in ZPL energies. Such large strain susceptibility was reported for the nitrogen antisite-vacancy pair \cite{liGiantShiftStrain2020}. A key question remains about the exact feature of the orbital structure, that enables the mechanical isolation. \\
On the experimental side, a major challenge is to improve spectral diffusion, where the optical transition frequencies vary over time due to a change in the local electric field of the emitter and thermal broadening. The linewidth of optical transitions was observed to be within the FTL for resonant excitation. However, spectral diffusion broadens the lines on long timescales, in particular, when off-resonant excitation is applied. Under off-resonant excitation the spectral linewidth typically broadens to many hundreds of GHz. The mitigation of spectral diffusion comes along with the question about the underlying processes. There is a strong difference between spectral diffusion under resonant and off-resonant excitation which needs further clarification. So far the more pronounced spectral diffusion under off-resonant excitation was assigned to the occupation of different excited states which have a stronger sensitivity to spectral diffusion. In general, spectral instabilities result in sever limitations for applications. As an example, the creation of indistinguishable single photons becomes very challenging. Furthermore, effects such as blinking \cite{sternSpectrallyResolvedPhotodynamics2019}, bleaching \cite{shotanPhotoinducedModificationSinglePhoton2016} or spectral diffusion \cite{liNonmagneticQuantumEmitters2017} limit the optical coherence properties \cite{spokoynyEffectSpectralDiffusion2020, sontheimerPhotodynamicsQuantumEmitters2017}. A detailed study of spectral diffusion was performed over a temperature range from 4 K to 300 K \cite{akbariTemperaturedependentSpectralEmission2021}. Three potential routes to overcome spectral diffusion are summarized in Figure \ref{fig:Diffusion}.

Spectral diffusion was suppressed by 45 \% by using a conductive substrate, thereby increasing the carrier density in the local environment of the defect center \cite{akbariTemperaturedependentSpectralEmission2021}. Figure \ref{fig:Diffusion} (A) shows the linewidth narrowing that occurs when the h-BN is place on a conducting substrate, such as ITO as compared to an insulating substrate, such as $SiO_2$. The spectral stability under resonant excitation was improved by repumping the system from the intermediate, nonradiative "dark" state. Optical co-excitation with a weak non-resonant laser at 500 nm reduces spectral blinking resulting in an order of magnitude improvement in the on-time under resonant excitation \cite{whiteOpticalRepumpingResonantly2020}, as depicted in Figure \ref{fig:Diffusion} (B). Yet another path to improve spectral stability is to utilize surface passivation, for example, by placing h-BN on a high-purity $Al_2O_3$ substrate \cite{liNonmagneticQuantumEmitters2017a}. Detrimental spectral diffusion was reduced by more than one order of magnitude from a spectral diffusion wavelength range of 5nm on $SiO_2$ substrate to about 0.4 nm when placed on uniform $Al_2O_3$, as shown in Figure \ref{fig:Diffusion} (C).

\section{Quantum photonics and plasmonics}

\begin{figure}
  \centering
  \includegraphics[width=15 cm]{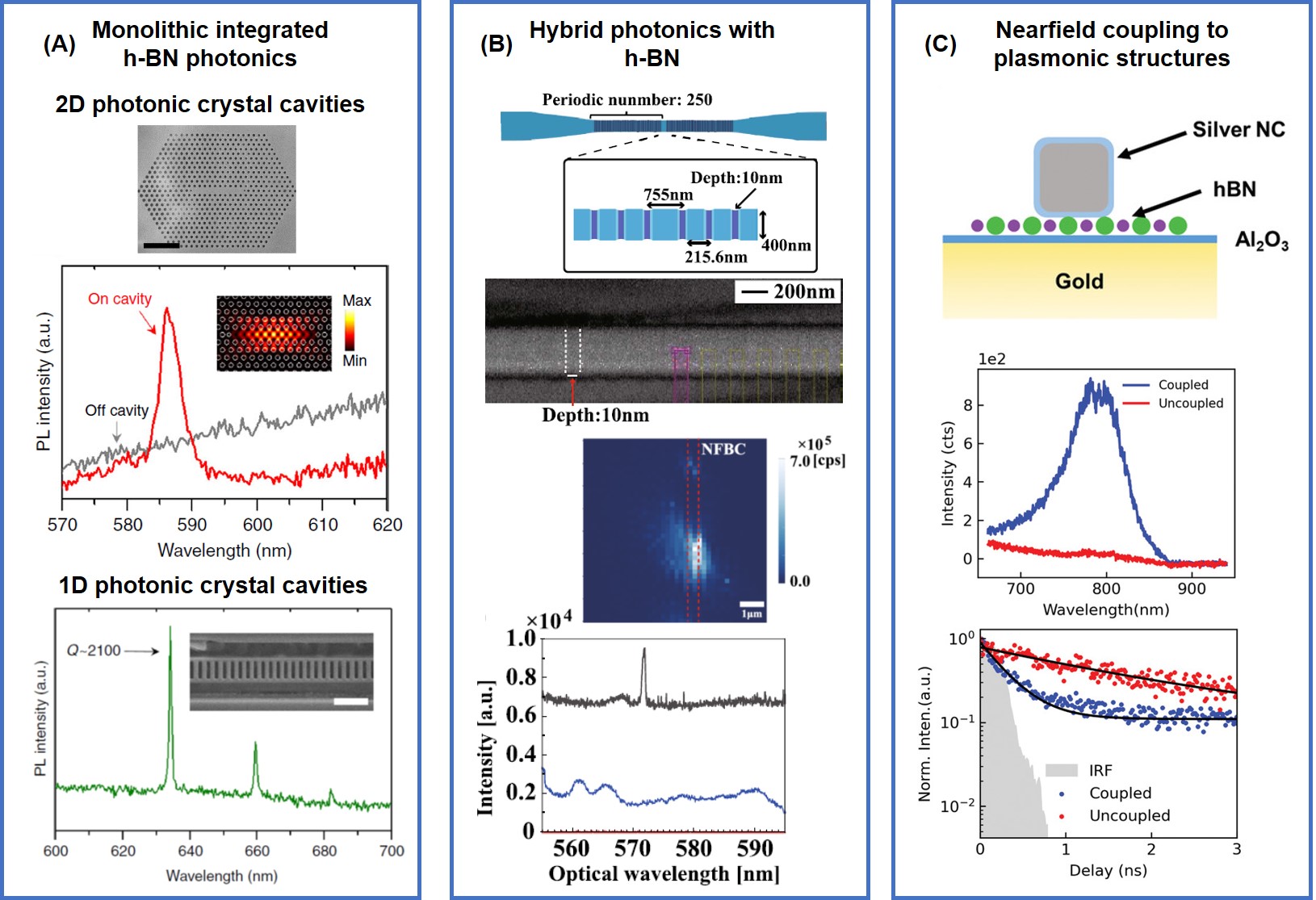}
  \caption{\textbf{h-BN based nanophotonics and plasmonics.} (A) Monolithic 1D and 2D h-BN photonic crystal cavities were fabricated reaching Q-factors above 2000 enabling Purcell factors beyond 100 under ideal coupling conditions. (B) Hybrid photonics with h-BN builds, for example, on advanced pick and place techniques to establish the hybrid platforms. As an example, nanofiber Bragg cavities are fabricated with Q-factors above 800 and postprocessed with h-BN nanoflakes (upper panel). The final hybrid device matches the fluorescence wavelength of an ensemble of defect centers resulting in light emission into the fiber (lower panel). (C) Few-layer hBN flakes integrated into plasmonic structures composed of ultraflat gold crystal and silver nanocubes (upper panel). Coupling $V_B^-$ emitters to the plasmonic gap cavity yields PL enhancement above 15 (central panel) and a PL decay rate increased by a factor of about 6 in resonance with the gap cavity mode (lower panel), corresponding to a Purcell factor $F_P$ of about 480. 
(A) Reproduced under terms of the CC-BY license \cite{kimPhotonicCrystalCavities2018a}. Copyright 2018, by Nature Communication published by Springer Nature. DOI:10.1038/s41467-018-05117-4.
(B) Reproduced under terms of the CC-BY license \cite{tashimaHybridDeviceHexagonal2022}. Copyright 2022, by Scientific Reports published by Springer Nature. DOI:10.1038/s41598-021-03703-z.
(C) Reprinted with permission from \cite{mendelsonCouplingSpinDefects2022}. Copyright 2021, by Advanced Materials published by Wiley-VCH GmbH. DOI: doi.org/10.1002/adma.202106046.}
  \label{fig:Nanophotonics}
\end{figure}

Recent research developments significantly improved spectral diffusion, as discussed in the previous section. Besides directly improving the spectral stability of the defect centers one could also broaden the homogeneous linewidth of the optical transition. One way to achieve a linebroadening is to utilize systems with large Purcell-factors. The Purcell-factor improves the ratio $\frac{\Gamma_{0}}{\Gamma_{inhom}}$ by lifetime-shortening therefore leading to a spectrally broadened line. Nanophotonics is a potential path to enhance the system efficiency and to improve the spectral properties by Purcell-assisted line-broadening. Mainly two directions are pursued to develop nanophotonics devices. \\
First, integrated photonics with h-BN as photonics material. Progress was achieved by developing suspended 1D and 2D photonic crystal cavities from h-BN with quality factors of more than 2000 \cite{kimPhotonicCrystalCavities2018a} as depicted in Figure \ref{fig:Nanophotonics} (A). Optimally coupled quantum emitters could reach a Purcell enhancement of about 110 for Q-values of 2100. Furthermore, 1D photonic crystal cavities, circular grating structures, waveguides and microrings \cite{frochPhotonicNanostructuresHexagonal2019} have been patterned on thin layers of exfoliated h-BN as well as ultrathin dielectric metalenses \cite{liuUltrathinVanWaals2018}. Designs for hybrid coupler promise Purcell enhancement of up to 285 with up to 89 \% collection efficiency \cite{yangHybridCouplerDirecting2021}. \\
Second, hybrid integration of defect centers in h-BN into photonic platforms. The hybrid approach takes advantage of the close proximity of the defect centers to the surface enabling strong, short-distance evanescent coupling. Highly-developed methods to realize hybrid devices are established. The pick-and-place technique enables high-precision integration of precharacterized h-BN into hybrid structures. Tools to precisely determine the emitters localization and orientation were developed, which enable optimized coupling and device assembly. The axial distance as well as the full dipolar orientation was determined for defect centers in multi-layer h-BN flakes by modification of the photonics density of states via phase-changing optical materials (VO$_2$) \cite{jhaNanoscaleAxialPosition2022}. Examples of hybrid quantum photonics include defect centers in h-BN optically coupled to nanofibers \cite{schellCouplingQuantumEmitters2017a}, $Si_3N_4$ microdisk optical resonators \cite{prosciaMicrocavitycoupledEmittersHexagonal2020}, waveguides  \cite{glushkovWaveguideBasedPlatformLargeFOV2019,kimIntegratedChipPlatform2019}, metallo-dielectric antennas \cite{liNearUnityLightCollection2019} and $Si_3N_4$ photonic crystal cavities \cite{frochCouplingHexagonalBoron2020}. Recently, ensemble defect centers in h-BN nanoflakes were coupled to nanofiber Bragg cavities by advanced pick and place technique \cite{tashimaHybridDeviceHexagonal2022}. The hybrid system reached Q-factors above 800 at a resonance wavelength of 573 nm which matches the transition wavelength of the defect centers as outlined in Figure \ref{fig:Nanophotonics} (B).
Growing uniform thin h-BN-films via chemical vapor deposition on SiO$_2$ microdisk cavities enables whispering gallery modes in the hybrid h-BN-SiO$_2$ microdisk with Q-factors larger $7.6 \times 10^5$ \cite{dasDemonstrationHybridHighQ2021}.
Furthermore, hybrid devices based on van der Waals heterostructures involving h-BN have been realized. Microdisk cavities fabricated from stacked h-BN / WS$_2$ / h-BN reached an enhanced emission intensity of the WS$_2$ trion by 2.9 times with a nonlinear power dependence when coupled to the whispering gallery modes \cite{renWhisperGalleryModes2018a}. MoSe$_2$-WSe$_2$ heterobilayers sandwiched between two h-BN slabs were used to fabricate waveguide-coupled disk resonators reaching maximum mode overlap and strong interlayer emission \cite{khelifaCouplingInterlayerExcitons2020}. \\
Plasmonics is an alternative route that gives access to large Purcell-factors \cite{baumbergExtremeNanophotonicsUltrathin2019}. Near-field coupling of a h-BN defect center to a resonant plasmonic nanoantenna revealed single h-BN emission sites of 45 nm size with lifetime shortening up to a factor of 2.2 \cite{palomboblascettaNanoscaleImagingControl2020}. $V_B^-$ emitters integrated into plasmonic cap cavities consisting of silver nanocubes and a gold mirror results in an enhancement in PL emission and lifetime reduction reaching Purcell-Factors of up to $F_P \approx 480$, as depicted in Figure \ref{fig:Nanophotonics} (C) \cite{mendelsonCouplingSpinDefects2022}. A PL enhancement up to 17-fold was reported for gold film surface plasmons \cite{gaoHighContrastPlasmonicEnhancedShallow2021}. Nanoassemblies of gold nanoshperes, controlled and positioned in the close proximity to quantum emitters in h-BN by AFM nanomanipulation were utilized to experimentally reach Purcell-factors of 3 with ideal theoretical Purcell factors of 1200 \cite{nguyenNanoassemblyQuantumEmitters2018}.

\begin{figure}
  \centering
  \includegraphics[width=16 cm]{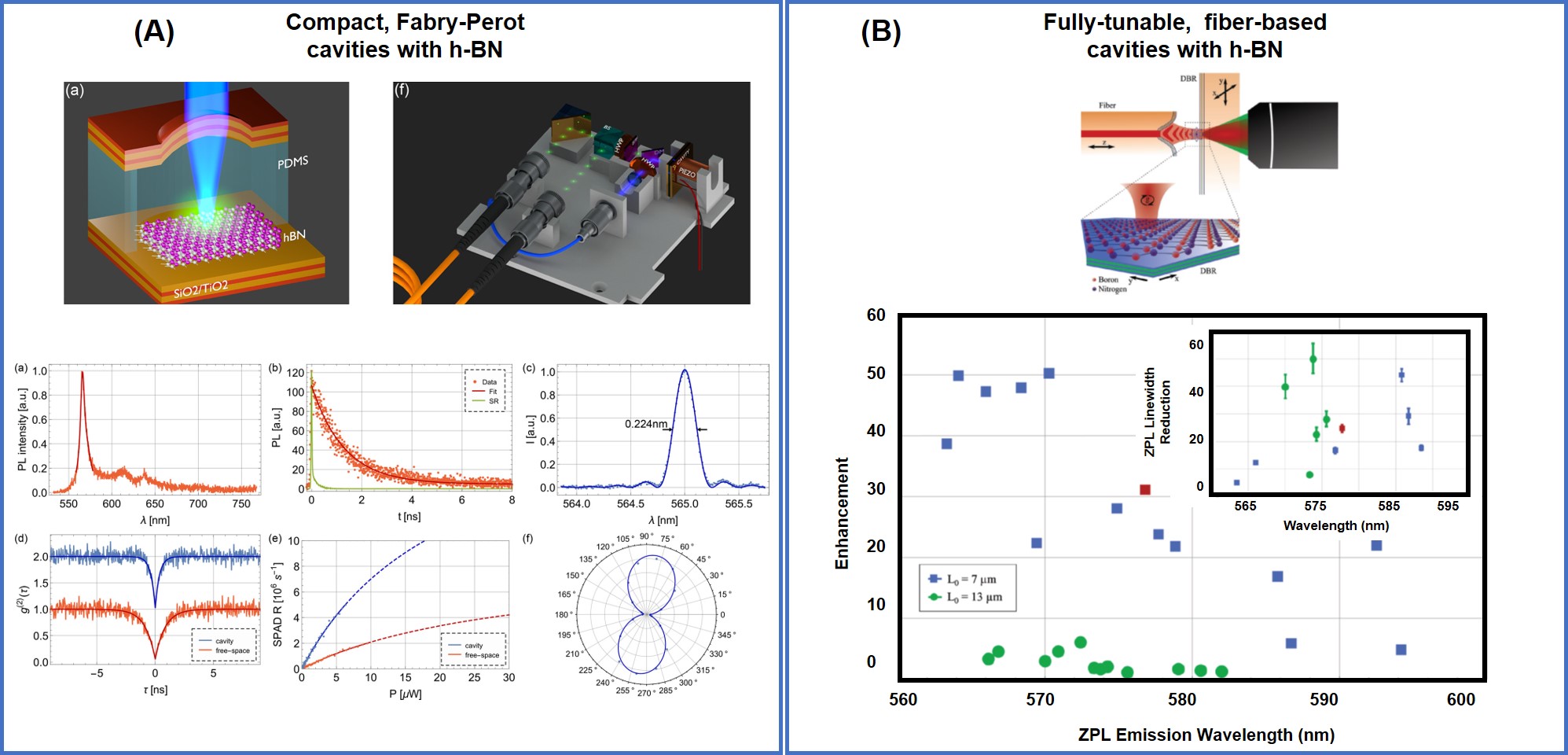}
  \caption{\textbf{Individual defect centers in h-BN integrated into optical resonators.} (A) Open Fabry-Perot resonators enable compact devices with high Purcell enhancement due to small mode volume on the order of $\lambda ^3$. The cavity narrows the linewidth from 5.76 nm to 0.224 nm. The lifetime is shortened by the Purcell effect from 837 ps to 366 ps leading to an improvement in the count rates, together with an enhanced collection efficiency. (B) Open fiber-based resonators enable low-loss, fully-tunable photonic interfaces for defect centers in h-BN. Up to 50-fold linewidth narrowing and signal enhancement was demonstrated for single defect centers. 
(A) Reprinted with permission from \cite{voglCompactCavityEnhancedSinglePhoton2019a}. Copyright 2019, American Chemical Society. DOI:10.1021/acsphotonics.9b00314.
(B) Reproduced under terms of the CC-BY license \cite{hausslerTunableFiberCavityEnhanced2021a}. Copyright 2021, by Advanced Optical Materials published by Wiley-VCH GmbH. DOI: doi.org/10.1002/adom.202002218.}
  \label{fig:Cavity}
\end{figure}

Cavity funneling improves the efficiency of both, excitation and emission paths, which also leads to improved spectral stability. Open resonators, in particular Fabry-Perot \cite{voglCompactCavityEnhancedSinglePhoton2019a} or fiber-based Fabry-Perot cavities \cite{hausslerTunableFiberCavityEnhanced2021a}, are low-loss, fully tunable alternatives to monolithic nanophotonics and plasmonics. Fabry-Perot resonators enable improved single-photon purity by strong signal enhancement and linewidth narrowing, as summarized in Figure \ref{fig:Cavity} (A). The linewidth narrows from a FWHM of 5.76 nm to 0.224 nm. The compact design with small mode volume ($\approx \lambda ^3$) enables robust operation, for example, in mobile applications. The in-situ tunability over one free-spectral range could enable to fabricate multiple, identical single photon sources \cite{voglCompactCavityEnhancedSinglePhoton2019a}. Figure \ref{fig:Cavity} (B) summarizes the results of coupling individual defect centers in h-BN to fully-tunable, fiber-based Fabry-Perot resonators. Cavity funneling manifests in strong linewidth narrowing by about a factor of 60 as well as signal enhancement by about 50 times. 

\section{Future perspectives and applications}

Optically active defect centers in wide band-gap materials with coherent optical transitions are a crucial constitute for future quantum optics applications. Isolated defect centers in h-BN with optical transition linewidths in the FTL even at room temperature could have large impact on coherent quantum technology featuring operation under ambient conditions. An open question remains about the exact orbital structure, which enables the mechanical isolation. Recently, simulations for out-of-plane defects, in particular group IV and V dimers, have been performed and discussed in the context of high-symmetry, high-spin defect centers\cite{bhangFirstPrinciplesPredictionsOutofPlane2021}. Applying similar modeling to the orbital structure of mechanically isolated defect centers could be a promising path forward to investigate the underlying decoupling mechanism. The apparent universality of the mechanical isolation mechanism, resulting in isolated emitters with a large range of ZPL transition wavelengths, enables applications in a broad range. As an example, experiments in nonlinear optics, optomechanics as well as quantum optics and photonics are discussed in this review. \\
Integrating single photon emitters, with spectral properties sensitive to mechanical strain, into h-BN optomechanical devices opens the door for spin-optomechanics with large zero point motion \cite{abdiSpinMechanicalSchemeColor2017}. Utilizing mechanically isolated emitters promise a new route towards spin-optomechanics with coherent optical transitions. A strong strain susceptibility for non-isolated emitters was demonstrated resulting in broadly distributed ZPL emission lines \cite{liGiantShiftStrain2020}. However, the strain susceptibility of isolated defect centers needs to be investigated and the mechanical isolation mechanism may indicate a weak strain susceptibility. \\
A main challenge is to coherently address and manipulate individual optical transitions and, ideally, coherently address associated spin states. Therefore, coherent optical control needs to be further developed. The main obstacle remains ongoing spectral diffusion. In this review, research activities directed to improve spectral diffusion are summarized. Furthermore, optical transitions can be spectrally broadened by employing the Purcell-effect, which improves the spectral stability. Promising routes with quantum photonics, plasmonics and open cavities are discussed. A look into the near future suggests that mechanically isolated defect centers with optical transition linewidths $\gamma _0$ as narrow as 60 MHz could be coupled to individual cavity modes. Estimating the coupling parameters based on state-of-the-art systems yields a cooperativity $C_0$ exceeding 100 at room temperature for the case of open cavities \cite{hausslerTunableFiberCavityEnhanced2021a}. Therefore, new directions towards solid-state based cavity quantum electrodynamics in the strong-coupling regime under ambient conditions become feasible. \\  
Further examples of quantum optics applications include quantum random number generation, as reported with \cite{whiteQuantumRandomNumber2021} and without \cite{hoeseSinglePhotonRandomness2022} beamsplitter with high entropy per raw bit. Having coherent optical transitions at hand brings applications based on indistinguishable single photon emission into reach, including quantum repeater, distributed quantum computing and quantum networks as well as photonics assisted quantum-metrology. \\
Summarizing, the mechanical isolation of defect centers in h-BN potentially enables coherent light-matter interaction at room temperature with the impact to revolutionize solid-state quantum optics and optomechanics by making sample cooling redundant enabling compact quantum devices operating under ambient conditions.

\medskip
%\textbf{Supporting Information} \par %Please delete the Suppporting Information statement if it is not applicable. Please supply Supporting Information in another file. Supporting information should not be provided in .tex format
%Supporting Information is available from the Wiley Online Library or from the author.

% Acknowledgements
\medskip
\textbf{Acknowledgements} \par %delete if not applicable))
A.K. acknowledges generous support provided by the Baden-Württemberg Stiftung in project 
Internationale Spitzenforschung and by IQst. A. K. acknowledges proof reading of the article, insightful discussions and experimental input from Michael Höse, Michael Koch and Vibhav Bharadwaj.
A. K. expresses his gratitude to Marcus Doherty and Igor Aharonovich for their theoretical and experimental input and insightful discussions.
A. K. offers his sincere thanks to Christoph Simon, Stephen Wein, Ken Sharman and Omid Gholami for stimulating discussions. 
A. K. acknowledges feedback on the manuscript from and insightful discussions with Paul Barclay, Dirk Englund, Gregory Fuchs, Nick Vamivakas, Isaac Luxmoore and Stefan Strauf.

% Harry Atwater, Ping Koy Lam
 
% Acknowledgements
\medskip
\textbf{Conflict of Interest} \par %delete if not applicable))
The authors declare no conflict of interest.

% References
\medskip

% Use the following code if you wish to generate your bibliography with BibTeX;
% replace the string "MSP-template" below with the name(s) of
% the BibTeX data base(s) you want to use.
% The resulting bibliography-output (the content of the .bbl file)
% must be pasted back into this file before submission.
% Please also include your BibTeX data base file(s) in your submission
% so that we can re-run BibTeX if necessary.
%
%\bibliographystyle{MSP}
%\bibliography{MSP-template}
%\bibliography{hBN.bbl}
\bibliographystyle{MSP}
\bibliography{hBN}

\end{document}